\documentclass[pra,twocolumn]{revtex4}
\usepackage{amsmath}
\usepackage{amssymb}
\usepackage{dsfont}
\usepackage{color}
\makeatletter
\usepackage{amssymb}
\usepackage{graphicx}
\usepackage{epstopdf}
\usepackage{epsfig}
\usepackage{subfigure}
\makeatother

\newcommand{\e}{\text{e}}

\DeclareMathAlphabet\mathbfcal{OMS}{cmsy}{b}{n}

\usepackage{cancel,ifthen}
\usepackage[normalem]{ulem}

\newcommand{\mje}[1]{{\color{black} #1}}

\newcommand{\comment}[2][NoInPuT]{\ifthenelse{\equal{#1}{NoInPuT}}{}{{\color{blue}\sout{#1}}}{\color{red} #2}}
\begin{document}

\title{Vortex patterns of atomic Bose-Einstein condensates in a density-dependent gauge potential}

\author{Matthew Edmonds}
\affiliation{Department of Physics \& Research and Education Center for Natural Sciences, Keio University, Hiyoshi 4-1-1, Yokohama, Kanagawa 223-8521, Japan}
\author{Muneto Nitta}
\affiliation{Department of Physics \& Research and Education Center for Natural Sciences, Keio University, Hiyoshi 4-1-1, Yokohama, Kanagawa 223-8521, Japan}

\date{\today{}}

\begin{abstract}\noindent
We theoretically examine the vortex states of a gas of trapped quasi-two-dimensional ultracold bosons subject to a density-dependent gauge potential, realizing an effective nonlinear rotation of the atomic condensate\mje{, which we also show is within the reach of current experimental techniques with ultracold atom experiments.} The nonlinear rotation has a two-fold effect; as well as distorting the shape of the condensate it also leads to an inhomogeneous vorticity resulting in novel morphological and topological states, including ring vortex arrangements that do not follow the standard Abrikosov result. The dynamics of trapped vortices are also explored, which differs from the case of \mje{rigid-body} rotation due to the absence of a global laboratory reference frame.
\end{abstract}
\maketitle
{\label{sec:int}\it Introduction. } Quantum vortices represent the fundamental excitations of superfluid systems, appearing in response to the rotation of atomic condensates. Unlike their classical counterparts a quantum vortex's rotational properties are more restricted, as their velocity field is quantized. The first generation of experiments used a laser to induce rotation by stirring the atomic cloud \cite{chevy_2000,raman_2001,abo_2001}. Vortices and in particular their interaction dynamics play a central role in condensed matter systems. A single vortex constitutes a core region where the density vanishes around which there is a circulation of the quantum fluid, consequentially there is a defect in the phase of the order parameter. Dimensionality plays a central role in the physics of vortices -- for two-dimensional quantum fluids experiments have demonstrated different topological phase transitions, prominent examples being the quantum spin Hall effect and the Berezinskii Kosterlitz Thouless transition, both of which have been realized with ultracold atoms \cite{beeler_2013,hadzibabic_2006}. In the three-dimensional context, more elaborate topological configurations are available, such as knots \cite{kawaguchi_2008,hall_2016,ollikainen_2019}, skyrmions \cite{kawakami_2012,choi_2012} and the related problem of engineering analogies of the magnetic monopole \cite{ruseckas_2005,ray_2014}.          

Ultracold atomic gases constitute exemplar physical systems for examining quantum mechanical phenomena, since experiments afford high controllability. Complementary to this, quantum gases represent dilute systems, which facilitates their accurate theoretical modelling via the celebrated Gross-Pitaevskii model and its numerous extensions \mje{\cite{dalfovo_1999,bloch_2008,kurn_2013,lahaye_2009}}. Vortices central role in superfluidity continues to attract theoretical and experimental interest in the macroscopic dynamics of these excitations. Early work focussed on studying the fundamental properties of the rotating system \cite{castin_1999,tsubota_2002,cozzini_2003}, while more recent work has \mje{centered} on understanding the effect of anisotropic trapping \cite{mcendoo}, vortex lattice \cite{riordan_2016a,riordan_2016b}, and chaotic \cite{zhang_2019a} dynamics. Focus has also been on the structure and dynamics of vortices in condensates at finite temperature including non-equilibrium effects \cite{liu_2018,kobayashi_2019}, multi-component systems which have been shown to possess a rich vortex physics \cite{kasamatsu_2003,eto_2011,kasamatsu_2016,mingarelli_2019,eto_2018,eto_2019} and the on-going quest to understand quantum turbulence \cite{tsatsos_2016,white_2014}. For a detailed description of the basic properties of rotating gases, the interested reader is directed to the review of Fetter \cite{fetter_2009}.        

The last decade has seen the experimental realization of artificial electromagnetism in quantum gases \cite{dalibard_2011,goldman_2014}. This provides an important new tool for accessing some of the paradigmatic effects associated with condensed matter physics, including simulating orbital magnetism \cite{lin_2009a,lin_2009b,lin_2011a,spielman_2009}. Ultracold atomic gases provide a platform for implementing not only these traditional manifestations of magnetism; but also for realizing forms of magnetism that are not naturally occurring, such as spin-orbit \cite{lin_2011b,campbell_2016} and spin-angular momentum coupling \cite{chen_2018,chen_2018b,zhang_2019} with bosons, synthetic dimensions \cite{celi_2014}, `knotted' gauge theories \cite{duncan_2019}, density-dependent magnetism \cite{edmonds_2013} \mje{and topological gauge theories \cite{rojas_2020}}. The realization of spin-orbit coupled bosons in-particular yields a new route to investigate the interplay of synthetic electromagnetism and rotation \cite{xu_2011,shi_2018,white_2017}. Induced electromagnetism in atomic gases typically produces a static gauge potential, here there is no feedback between the light and the matter-wave. To address this, several proposals have been put forward \cite{banerjee_2012,zohar_2013,tagliacozzo_213} to instead simulate $\textit{dynamical}$ gauge potentials. Very recently the first generation of experiments to realize density-dependent magnetism have appeared, for bosons \cite{clark_2018} and fermions \cite{gorg_2019} trapped in two-dimensional optical lattices, as well in a system of Rydberg atoms \cite{lienhard_2020}. 

\mje{In this Rapid Communication we examine the physics of a gas of confined two-dimensional bosons experiencing a density-dependent gauge potential that constitutes an effective nonlinear rotation of the atomic system, which we further illustrate is within the reach of the current generation of experiments. Our work builds on existing phenomenology while exploring a qualitatively novel form of synthetic magnetism that is presently under active experimental investigation \cite{clark_2018,gorg_2019,lienhard_2020}. We reveal the unusual phenomenology of this system, including vortex ring arrangements that violate the famous Abrikosov result and novel dynamics associated with the time-dependent nature of the underlying synthetic gauge potential.}

{\label{sec:model}\it Theoretical model. } We consider a system comprising $N$ two-level atoms coupled via a coherent light-matter interaction, forming a Bose-Einstein condensate. The Hamiltonian describing our setup can be written \mje{in the rotating wave approximation} as \mje{\cite{dalibard_2011}}
\begin{equation}\label{eqn:ham}
\hat{H}=\bigg(\frac{\hat{\bf p}^2}{2m}+V_{\rm ext}({\bf r})\bigg)\otimes\mathds{1}+\hat{H}_{\rm int}({\bf r})+\hat{U}_{\rm MF}
\end{equation}
where the light-matter interaction is defined \mje{\cite{goldman_2014}}
\begin{equation}\label{eqn:hlm}
\hat{U}_{\rm MF}=\frac{\hbar\Omega_r}{2}\left(\begin{array}{cc}\cos\theta & e^{-i\phi}\sin\theta \\e^{i\phi}\sin\theta & -\cos\theta\end{array}\right).
\end{equation}
The other quantities appearing in Eq.~\eqref{eqn:ham} are the external trapping potential $V_{\rm ext}({\bf r})$, and the mean-field interactions $\hat{H}_{\rm int}=(1/2)\text{diag}[\Delta_{1},\Delta_2]$, where $\Delta_{j}=g_{jj}n_j+g_{jk}n_k$ and  $g_{jk}=4\pi\hbar^2 a_{jk}/m$ defines the scattering parameter between atoms in the internal states $j$ and $k$, while $n_j=|\psi_j({\bf r})|^2$ defines the population density of state $j$. Meanwhile the harmonic trapping potential is given by $V_{\rm ext}({\bf r})=m(\omega_{x}^{2}x^2+\omega_{y}^{2}y^2+\omega_{z}^{2}z^2)/2$, where $\omega_j$ defines the strength of the trapping potential in each coordinate direction. To build an interacting gauge theory, we construct interacting dressed states using perturbation theory which is valid when the mean-field interactions are weak compared to the strength of the light-matter coupling. Denoting the (unperturbed) eigenstates of $\hat{U}_{\rm MF}$ (Eq.~\eqref{eqn:hlm}) as $|\pm\rangle$, the interacting dressed state basis can be written as \mje{\cite{edmonds_2013}}
\begin{equation}\label{eqn:pb}
|\psi_\pm\rangle=|\pm\rangle\pm\frac{\Delta_{d}}{\hbar\Omega_r}|\mp\rangle.
\end{equation}
The dressed mean-field detuning is $\Delta_{\rm d}=\sin\frac{\theta}{2}\cos\frac{\theta}{2}(\Delta_1-\Delta_2)/2$ \mje{and in what follows we use the adiabatic condition which assumes that only one of the dressed states $|\pm\rangle$ is occupied, such that $\hbar\Omega_r\gg E_{R}$, where $E_R=p_{\rm R}^{2}/2m$ and $p_{\rm R}=\hbar k$ define the recoil energy and momentum respectively \cite{cheneau_2008}. Using physical parameters appropriate for Sr, one obtains $E_{\rm R}/\hbar\simeq 58$kHz, while the current generation of experiments can achieve coupling strengths of $\Omega_r\simeq 2\pi\times 100$kHz \cite{zhang_2019}, giving $\hbar\Omega_R/E_R\simeq 11$.} To build the interacting gauge theory, we construct a state vector from the two perturbed dressed states defined by Eq.~\eqref{eqn:pb}. The qualitative features do not depend on this choice, here we project the atoms motion into the $+$ state. Then, the effective Hamiltonian becomes
\begin{equation}\label{eqn:he}
\hat{H}_{+}=\frac{({\bf p}-{\bf A}_{+})^2}{2m}+W_{+}+\frac{\hbar\Omega_r}{2}+\Delta_{+}+V_{\rm ext}({\bf r})
\end{equation}  
where the two \mje{geometric potentials} that arise due to the adiabatic motion of the atoms are given by ${\bf A}_{+}=i\hbar\langle\psi_{+}|\nabla\psi_{+}\rangle$ and $W_{+}=\hbar^2|\langle\psi_+|\nabla\psi_-\rangle|^2/2m$ respectively, while the dressed mean-field interactions are defined by $\Delta_+=(\Delta_1\cos^2\frac{\theta}{2}+\Delta_2\sin^2\frac{\theta}{2})/2$. These \mje{individual} terms can be shown to be defined as
\begin{figure}
\includegraphics[width=\columnwidth]{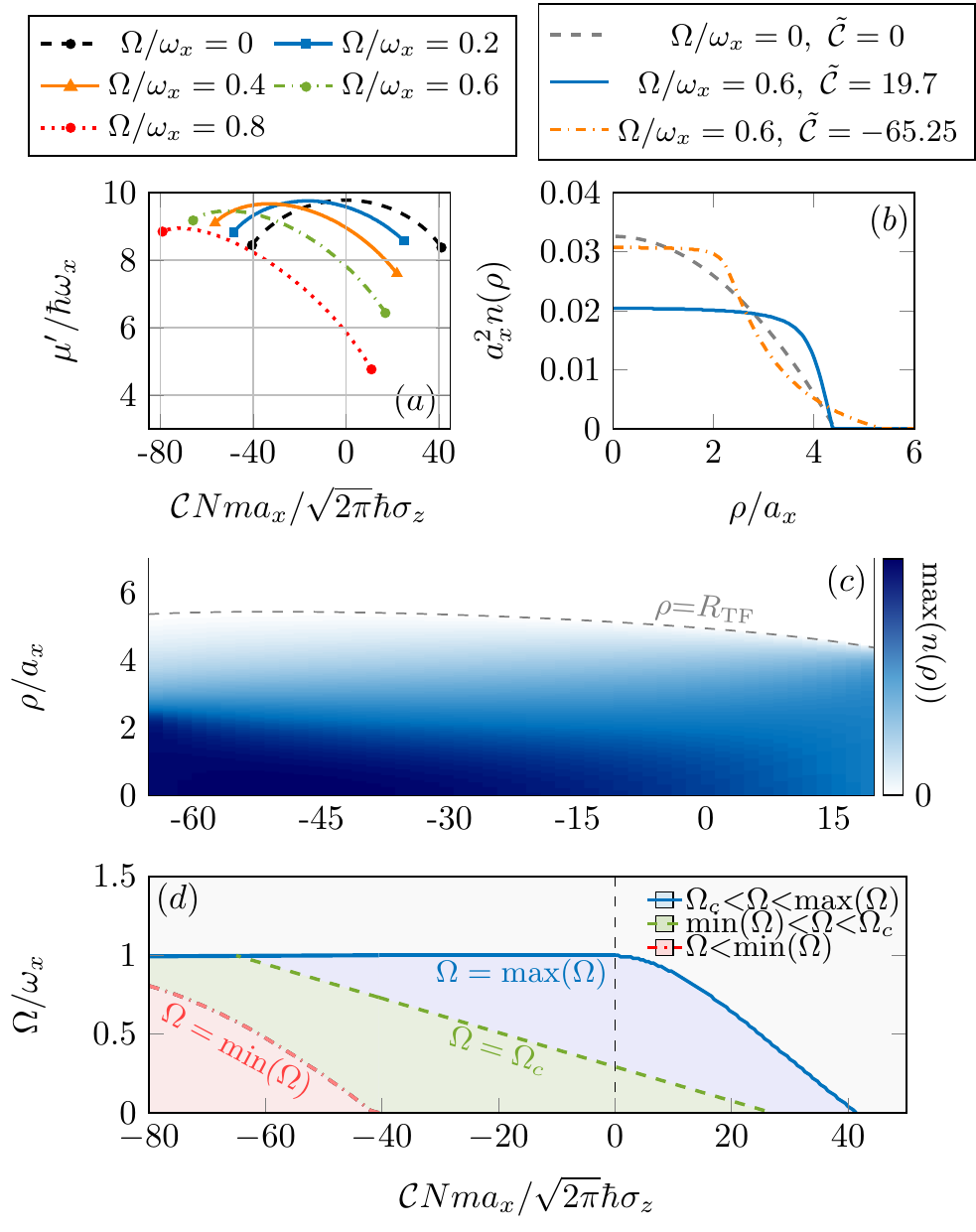}
\caption{\label{fig:omega}(color online) Rotating Thomas-Fermi solutions. (a) shows the solutions to Eq.~\eqref{eqn:tf} for different fixed values of the \mje{rigid-body} rotation frequency, $\Omega/\omega_x$. (b) compares the different density distributions that are obtained with $\Omega/\omega_x=0.6$ from varying $\mathcal{C}$, while (c) shows a heatmap of the radial density. Panel (d) displays the boundaries \mje{between regions where vortices are energetically favorable (blue region, blue solid line), unfavorable (green and red regions, green dashed and red dot-dashed) as well as where the Thomas-Fermi approximation, Eq.~\eqref{eqn:tf} breaks down (red dot-dashed).} Throughout we have used $ma_xg_{\rm eff}N/\sqrt{2\pi}\sigma_z\hbar^2=300$.}
\end{figure}

\begin{subequations}
\begin{align}\label{eqn:avec}
{\bf A}_+=&-\frac{\hbar}{2}(1-\cos\theta)\nabla\phi+\frac{\Delta_{\rm d}}{\Omega_r}\nabla\phi\sin\theta,\\ \nonumber
W_+=&\frac{\hbar^2}{8m}(\nabla\theta)^2+\frac{\hbar}{8m}\sin^2\theta(\nabla\phi)^2\\+&\frac{\hbar}{2m}\frac{\Delta_{\rm d}}{\Omega_r}\sin\theta\cos\theta(\nabla\phi)^2-\hbar\nabla\theta\cdot\nabla\frac{\Delta_{\rm d}}{\Omega_r}.\label{eqn:wsca}
\end{align}
\end{subequations}
To obtain a mean-field equation of motion for the atoms, we extremize the energy functional $\mathcal{E}=\langle\psi_+|\hat{H}_+|\psi_+\rangle$, leading to the following generalized Gross-Pitaevskii equation \mje{(valid specifically at $T=0$)}\cite{butera_2017,butera_2015,butera_2016}
\begin{align}\nonumber
i\hbar&\frac{\partial\psi_+}{\partial t}{=}\bigg[\frac{({\bf p}-{\bf A}_{+})^2}{2m}+W_{+}{+}{\bf a}_1\cdot{\bf j}{+}\frac{\hbar\Omega_r}{2}{+}2\Delta_+{+}V({\bf r})\bigg]\psi_+\\&+\bigg[n_+\bigg(\frac{\partial W_+}{\partial \psi^{*}_{+}}-\nabla\cdot\frac{\partial W_+}{\partial\nabla\psi^{*}_{+}}\bigg)-\frac{\partial W_+}{\partial\nabla\psi^{*}_{+}}\cdot\nabla n_+\bigg].\label{eqn:gpe}
\end{align}
Where the coupling to the gauge field is defined as ${\bf a}_1=\nabla\phi\Delta_{d}\sin\theta/n_{+}\Omega_r$. Then, the density-dependent dressed basis, Eq.~\eqref{eqn:pb} gives rise to additional terms in the generalized Gross-Pitaevskii equation for atoms in the $\psi_+$ state, including the current nonlinearity ${\bf j}$ which appears as
\begin{equation}\label{eqn:j}
{\bf j}=\frac{\hbar}{2mi}\bigg[\psi_+\bigg(\nabla+\frac{i}{\hbar}{\bf A}_+\bigg)\psi^{*}_+-\psi^{*}_{+}\bigg(\nabla-\frac{i}{\hbar}{\bf A}_+\bigg)\psi_+\bigg]
\end{equation}
The current nonlinearity Eq.~\eqref{eqn:j} gives rise to novel topological states; in-particular proposals to simulate exotic spacetime geometries \cite{butera_2019} in the three-dimensional context. In the one-dimensional limit the theory violates Kohn's theorem \cite{edmonds_2015,zheng_2015}, and exact chiral soliton solutions \cite{dingwall_2018} can also be constructed in this limit, which have recently been shown to constitute quantum time crystals \cite{ohberg_2019}. \mje{Complementary to the continuum method, we also note the lattice based approaches, where the density-dependent gauge potential enters via the appropriate Peierls phase \cite{keilmann_2011,greschner_2014,greschner_2015}.} 
We identify two small parameters: $\theta=\Omega_r/\Delta$, the ratio of the Rabi frequency to the detuning, and $\varepsilon=n(g_{11}-g_{12})/4\hbar\Delta$ that encompasses the collisional and coherent interactions. After expanding Eqs.~\eqref{eqn:avec}-\eqref{eqn:wsca} to linear order in $\varepsilon$ and $\theta$ we obtain simplified expressions for the vector potential \mje{${\bf A}_+=-\hbar\theta^2\nabla\phi[1-4\varepsilon]/4$, and $W_+$, the scalar potential}
\mje{\begin{equation}
W_{+}{=}\frac{\hbar^2}{2}\bigg[\frac{(\nabla\theta)^2[1{-}4\varepsilon]{+}\theta^2(\nabla\phi)^2[1{+}4\varepsilon]}{4m}{-}\nabla\theta^2\cdot\nabla\varepsilon\bigg].\label{eqn:wsca2}
\end{equation}}
To build the interacting gauge theory, we choose a laser beam carrying \mje{$\ell=+1$} units of angular momentum such that the phase $\phi=\ell\varphi$, where $\varphi$ is the polar angle \cite{juzeliunas_2005}. The spatial profile meanwhile satisfies $\theta=\theta_0 r$, where $r$ is the radial distance. \mje{Our choice here is an experimentally motivated one, since Laguerre-Gaussian laser light carrying $\ell=+1$ units of angular momentum with a cylindrically varying intensity profile has recently been used in the experimental demonstration of spin-angular-momentum coupled Bose-Einstein condensates \cite{chen_2018,chen_2018b}.} Then, using Eqs.~\eqref{eqn:gpe}, \eqref{eqn:j} along with the simplified expressions for ${\bf A}_+$ and $W_+$ \mje{(Eq.~\eqref{eqn:wsca2})}, the equation of motion for the condensate becomes
\begin{equation}\label{eqn:gpe_rot}
i\hbar\frac{\partial\psi}{\partial t}{=}\bigg[{-}\frac{\hbar^2}{2m}\nabla^2{+}V_{\rm ext}({\bf r}){-}\Omega({\bf r},t)\hat{L}_z{+}g_{\rm eff}n({\bf r})\bigg]\psi,
\end{equation}
here the angular momentum operator is defined as $\hat{L}_z=-i\hbar\partial/\partial\varphi$. Then we define the density-dependent rotation frequency as $\Omega({\bf r},t)=\Omega+\mathcal{C}n({\bf r},t)$ where the strength of the nonlinear \mje{rotation} term is $\mathcal{C}=\theta_{0}^2(g_{11}-g_{12})/(2m\Omega_r)$, while the effective scattering parameter is given by \mje{$g_{\rm eff}=g_{11}+\theta_{0}^2\hbar(g_{11}-g_{12})/m\Omega_r$}. As we are interested in the quasi-two-dimensional situation, we assume a pancake geometry for the atomic cloud ($\omega_z\gg\omega_{x,y}$), which allows us to factorize the atomic wave function as $\psi({\bf r},t)=\psi(x,y,t)e^{-z^2/2\sigma_{z}^{2}}/(\sqrt[4]{\pi\sigma_{z}^{2}})$, which has the effect of rescaling the two nonlinear terms $g_{\rm eff}$ and $\mathcal{C}$ by $1/\sqrt{2\pi}\sigma_z$.   

\begin{figure}[t]
\includegraphics[width=\columnwidth]{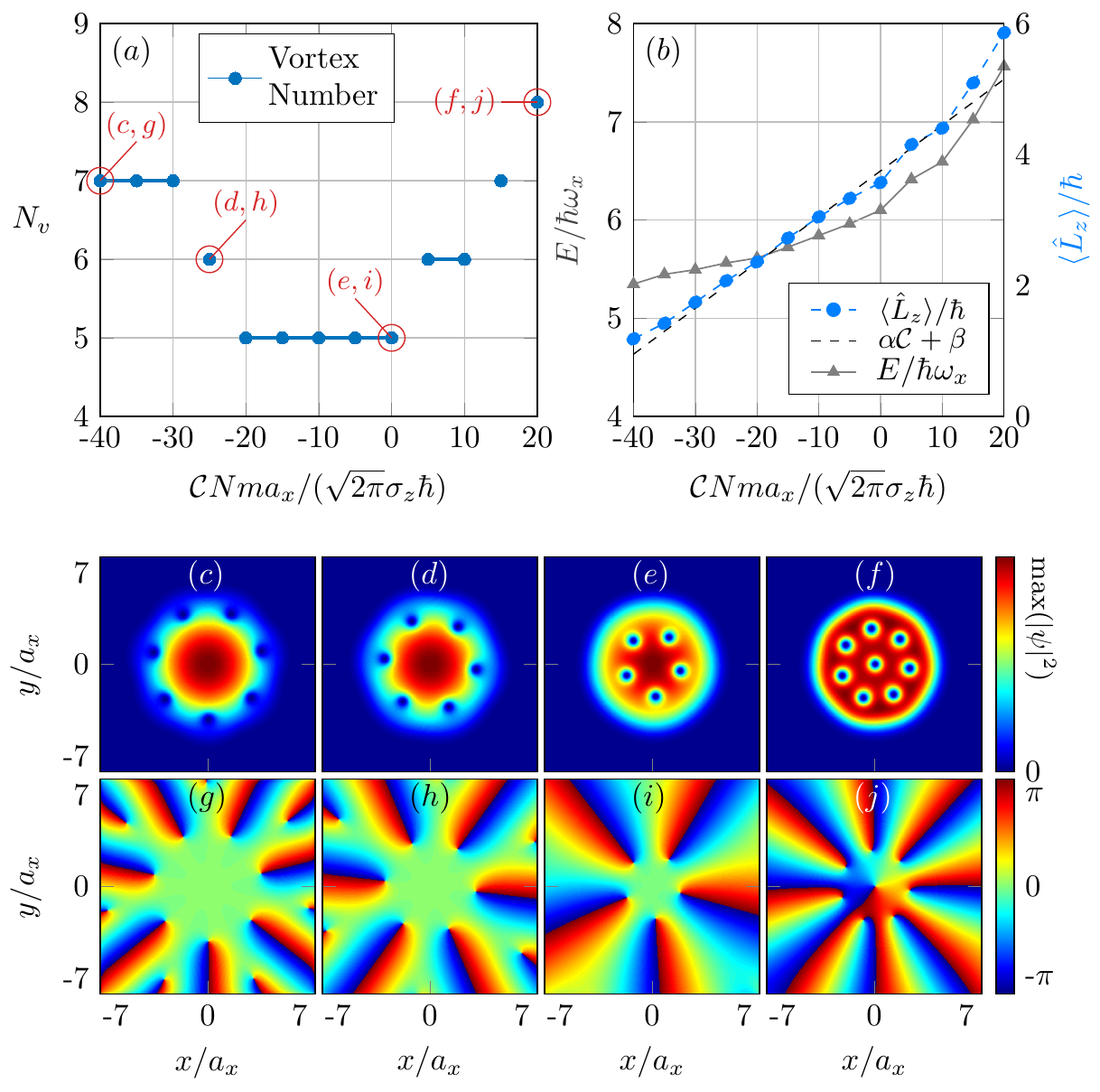}
\caption{\label{fig:iso}(color online) Vortex ground states. Panel (a) shows the number of vortices as a function of $\tilde{C}$, while (b) presents the energy and angular momentum. The corresponding ground state densities and phase profiles are shown in (c)-(j).}
\end{figure}
\mje{To realize the model of Eq.~\eqref{eqn:gpe_rot} we require that the motion of the atoms be adiabatic in one of the dressed states, as well as satisfying the perturbative requirement that underpins Eq.~\eqref{eqn:pb}. For these purposes the alkali-earth atoms represent a promising candidate, since the excited states of these atoms possess lifetimes of the order of seconds, such as the transition $^{3}S_{0}\leftrightarrow {^{1}}P_{1}$ in Sr, which could be used in a future experiment \cite{ye_2008}. Accompanying this one also needs in be in a regime where the perturbative assumption underlying Eq.~\eqref{eqn:pb} is valid. This can be achieved with the aid of optical Feshbach resonances for alkali-earth atoms \cite{enomoto_2008}. Again using physical parameters for Sr, a difference of scattering lengths $a_{11}-a_{12}=10$nm with an atomic density $n=10^{15}$cm$^{-3}$ and $\Omega_r=2\pi\times 100$kHz, one finds $\varepsilon\simeq 4\times 10^{-2}$, which is small enough  to justify perturbation theory but large enough to potentially observe the effects described in this work.}

{\it Ground states and vorticity. } To gain an understanding of the basic physics associated with a trapped condensate under nonlinear rotation, we begin with the energy functional associated with Eq.~\eqref{eqn:gpe_rot} which can be written in the hydrodynamic prescription using the Madelung transformation $\psi=\sqrt{n}e^{i\phi}$ where $n\equiv n({\boldsymbol{\rho}},t)$ is the quasi-two-dimensional density while $\phi\equiv\phi({\boldsymbol{\rho}},t)$ is the corresponding phase, then after dropping the quantum pressure term (valid for $ma_xg_{\rm eff}/\sqrt{2\pi}\sigma_z\hbar^2\gg1$) we obtain
\begin{equation}\label{eqn:engpe}
E{=}\int d^2{\bf r}n\bigg[\frac{m{\bf v}^2}{2}{+}\frac{i\hbar \nabla n}{2mn}\cdot{\mathbfcal{A}}{-}\frac{\mathbfcal{A}^2}{2m}{+}\frac{g_{\rm eff}n}{2}{+}V_{\rm 2D}\bigg],
\end{equation} 
here ${\bf v}=(\hbar\nabla\phi-\mathbfcal{A})/m$ defines the kinetic velocity with $\mathbfcal{A}=m\boldsymbol{\Omega}\times\boldsymbol{\rho}$, $\boldsymbol{\Omega}=\hat{e}_z(\Omega+\mathcal{C}n/2)$, and $V_{\rm 2D}(x,y)=m(\omega_{x}^{2}x^2+\omega_{y}^{2}y^2)/2$. The energy defined by Eq.~\eqref{eqn:engpe} can be minimized to obtain the superfluid velocity ${\bf v}_{\rm sf}=\boldsymbol{\Omega}\times\boldsymbol{\rho}$ 
which couples to both the \mje{rigid-body} rotation \mje{strength} $\Omega$ and the density $n(\rho)$ of the gas. Using Eq.~\eqref{eqn:engpe} and the expression for the superfluid velocity, we obtain a generalized Thomas-Fermi distribution in the rotating frame as
\begin{equation}\label{eqn:tf}
V_{\rm 2D}(x,y)-\frac{m}{2}\rho^2\bigg(\Omega^2+2\Omega\mathcal{C}n+\frac{3}{4}\mathcal{C}^2n^2\bigg)+g_{\rm eff}n=\mu',
\end{equation}
where $\rho^2=x^2+y^2$ and $\mu'$ is the chemical potential in the rotating frame, and the radius of the cloud is defined as $R^{2}_{x,y}=2\mu'/m(\omega_{x,y}^2-\Omega^{2})$, while $n(\rho)$ is subject to the normalization $\int d^2{\bf\rho}\ n({\bf\rho})=N$. 
\begin{figure}[b]
\includegraphics[width=\columnwidth]{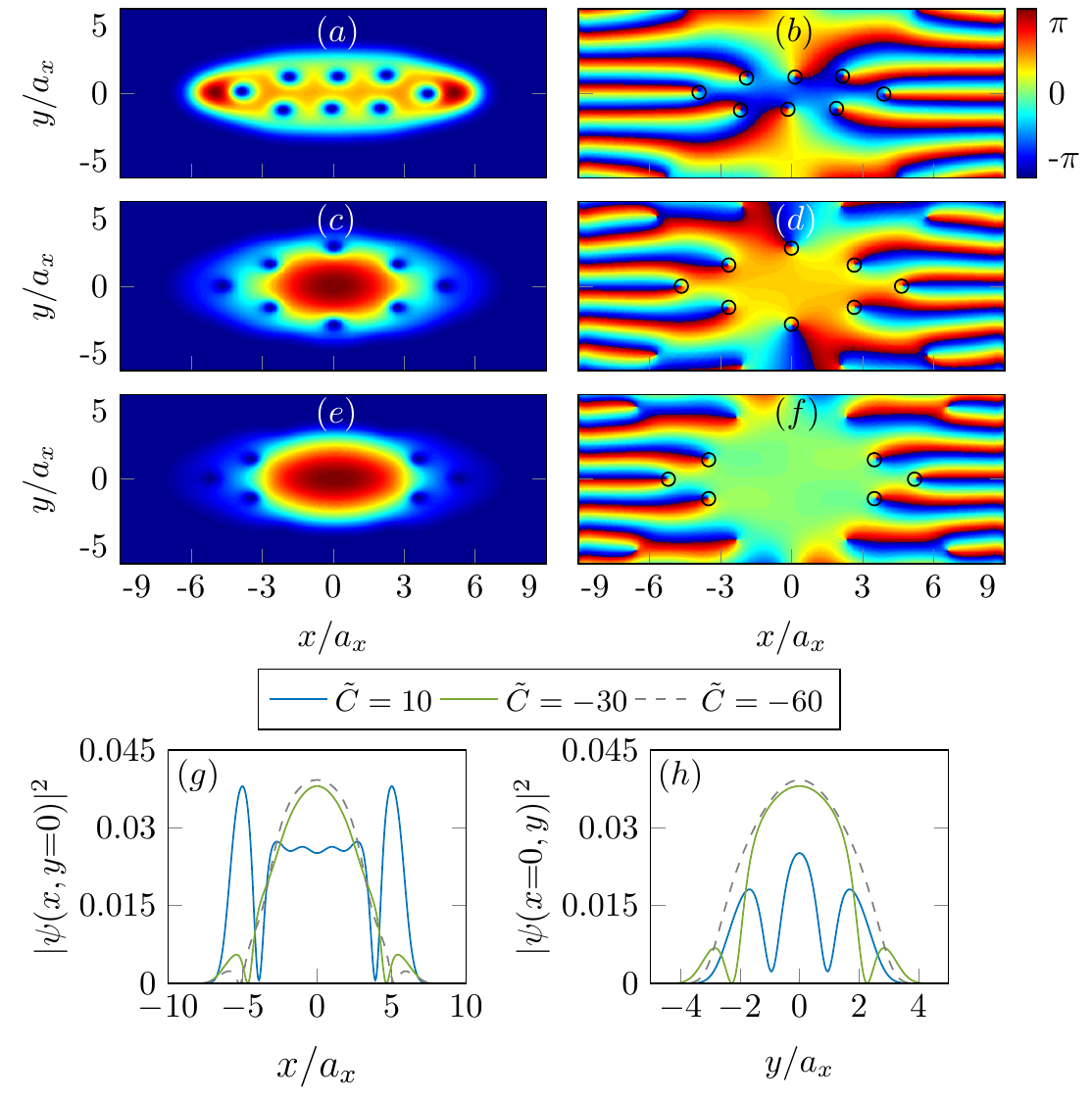}
\caption{\label{fig:ell}(color online) Elliptical trap ground states. Panels (a)-(f) show ground state density and phase profiles for $\omega_y/\omega_x=1.5$. Each row corresponds to $\tilde{C}=10,-30,-60$. The vortices are highlighted by black circles in the phase, right column. Panels (g) and (h) show cuts of the density along the coordinate axis ($x,y=0$) corresponding to the three nonlinear rotation strengths.}
\end{figure}
Figure \ref{fig:omega} (a) and (b) compute the chemical potential $\mu'$ and density $n({\bf\rho})$. Panel (a) shows that the solutions for $\mu'$ exist on finite regions, between a minimum and a maximum $\mathcal{C}$. The maximum value of $\mathcal{C}$ in each case corresponds to the point where the gas locally exceeds the trapping frequency, while the minimum $\mathcal{C}$ corresponds instead to the point when the approximations leading to Eq.~\eqref{eqn:tf} \mje{break down}. The solid blue curve in panel (b) shows an example for large positive $\mathcal{C}$, where the density profile exhibits a large plateau region, reminiscent of a quantum droplet \cite{tengstrand_2019}. For large negative $\mathcal{C}$, the tails of the distribution appear to decay as they approach the edge of the cloud, with a small central plateau region. The radial density $n(\rho)$ is plotted as a function of $\mathcal{C}$ in panel (c), for $\Omega/\omega_x=0.6$. The final panel (d) of Fig.~\ref{fig:omega} computes the boundaries that the solutions occupy in the $(\Omega,\mathcal{C})$ parameter space. Accompanying this is the critical rotation frequency $\Omega_c$ at which it is energetically favorable for a vortex to enter the cloud \cite{lundh_1997,butera_2016}, calculated from $\Omega_c=(\langle E_v\rangle-C\langle n\hat{L}_z\rangle/2)/\langle\hat{L}_z\rangle$, (green dashed line) using the non-rotating ($\Omega({\bf r},t)=0$) \mje{Thomas-Fermi solution $n=(\mu'-V_{\rm 2D}(x,y))/g_{\rm 2D}$, with $\mu'=\hbar\omega_x(g_{\rm 2D}Nm\omega_y/(\pi\hbar^2\omega_x))^{1/2}$ }. 
We define the average vorticity as $\boldsymbol{\omega}_v=\nabla\times{\bf v}$, which can be calculated from $\boldsymbol{v}=\Omega({\bf r},t)\hat{e}_z\times\boldsymbol{\rho}$ giving
\begin{equation}\label{eqn:wv}
\boldsymbol{\omega}_v(\rho,t)=\hat{e}_z\bigg[2\Omega(\rho,t)+\rho\frac{\partial\Omega(\rho,t)}{\partial\rho}\bigg],
\end{equation}
and the synthetic magnetic field is related to Eq.~\eqref{eqn:wv} since ${\bf B}(\rho,t)=m\boldsymbol{\omega}_v(\rho,t)$. \mje{Here it is worth considering a broader picture of the physical consequences the density-dependent rotation. Rigid-body rotation constitutes a global gauge symmetry, leading to the manifestation of phase defects, vortices in atomic Bose-Einstein condensates \cite{chevy_2000,raman_2001,abo_2001}. In contrast, spatially varying synthetic magnetic fields possess a local gauge symmetry, which can lead to the generation of more elaborate topological excitations, such knots, skyrmions and also monopoles. It is also worth noting that the study of synthetic gauge theories is also being pursued in other fields, such as the growing field of emergent electromagnetism in condensed matter systems \cite{benton_2012,sibille_2018}.}
\begin{figure}[t]
\includegraphics[width=.9\columnwidth]{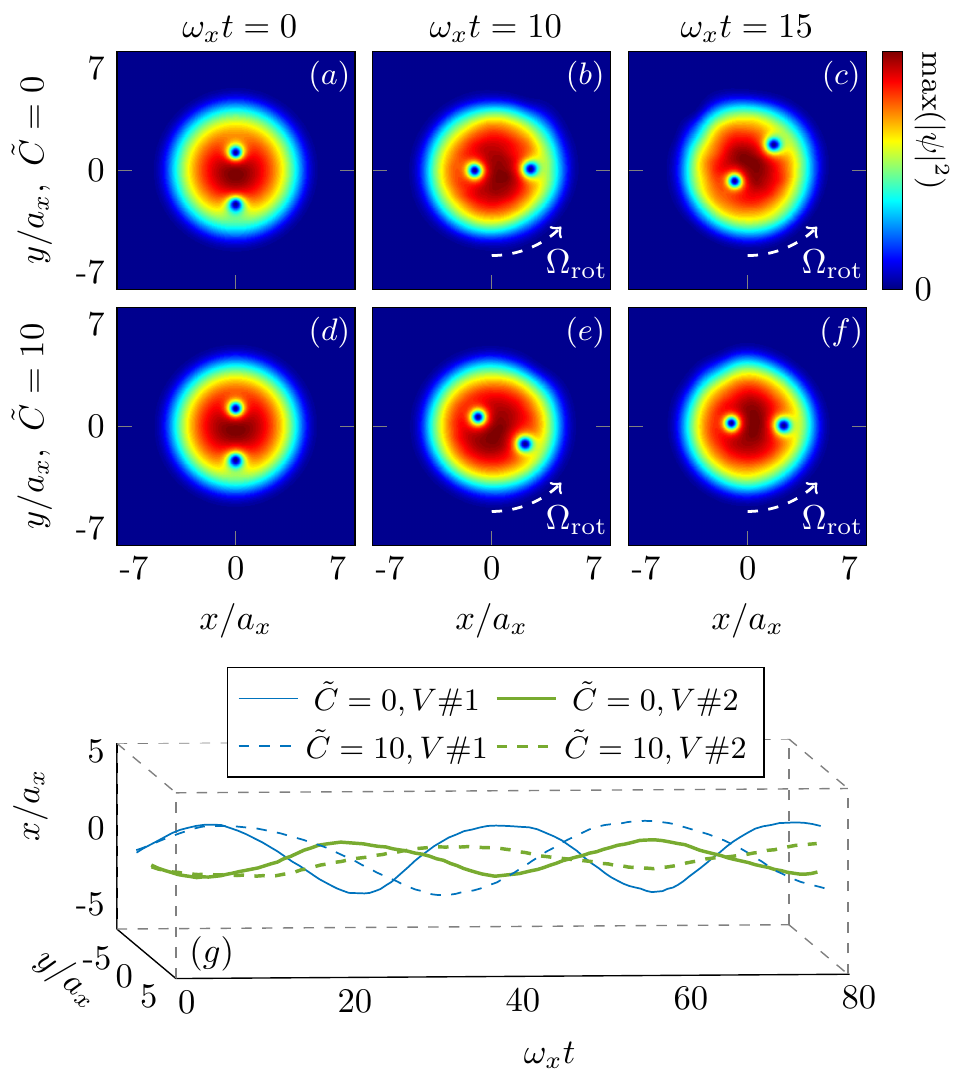}
\caption{\label{fig:dyn}(color online) Asymmetric vortex dynamics. Panels (a-f) show snapshots of a pair of vortices positions for two different values of $\tilde{C}$ for $\omega_xt=0,10,15$. (g) shows the time evolution of each pairs trajectory in $x,y$ as a function of $\omega_x t$. The anti-clockwise rotation direction is indicated by $\Omega_{\rm rot}$.}
\end{figure}

{\label{sec:numerics}\it Numerical simulations. } 
Figure \ref{fig:iso} presents calculations of the ground states of Eq.~\eqref{eqn:gpe_rot} as a function of the strength of the nonlinear rotation $\mathcal{C}$. Throughout we take the van der Waals strength as $ma_xg_{\rm eff}N/(\sqrt{2\pi}\sigma_z\hbar^2)=300$ \mje{and we define the dimensionless nonlinear rotation strength as $\tilde{\mathcal{C}}=\mathcal{C}Nma_x/(\sqrt{2\pi}\hbar\sigma_z)$}, while the \mje{rigid-body} rotation strength $\Omega/\omega_x=0.6$, and the trap anisotropy is $\omega_y/\omega_x=1.01$. Panel (a) shows the number of vortices as the strength of $\mathcal{C}$ is changed. The labels correspond to the panels (c) to (j) showing the atomic density $n(\rho)$ and phase $\phi(\rho)$ of the ground state in each case. 
For large positive $\mathcal{C}$, an unusual cylindrical density is observed (panel f), corroborating the analytical prediction, Fig.~\ref{fig:omega}(b) which in the presence of vortices resembles a $\textit{Renkon}$ lotus root. For negative $\mathcal{C}$, the triangular lattice is no-longer observed. Instead the vortices arrange into rings with increasing radius as $\mathcal{C}$ is decreased (for larger $\Omega$, concentric vortex ring arrangements are observed when $\tilde{C}\ll0$). The localization of the vortices can be interpreted from Eq.~\eqref{eqn:wv} due to the vorticity being maximum in the center of the cloud when $\tilde{C}>0$, while for $\tilde{C}<0$ it is maximal at the edges of the cloud, leading to ring arrangements. Panel (b) shows the energy $E$ (gray triangles) and angular momentum $\langle\hat{L}_z\rangle$ (solid blue circles) along with a linear fit for the angular momentum (see also Ref.~\cite{correggi_2019} concerning vortex patterns in a related anyonic model).

Next we explore the effect of a general elliptical trapping potential ($\omega_y\gg\omega_x$) on the ground state in fig.~\ref{fig:ell}. The parameters are chosen such that $\omega_y/\omega_x=1.5$ and $\Omega/\omega_x=0.8$. Panels (a)-(f) show the density and phase for three individual values of the strength of the nonlinear rotation. Panels (g) and (h) show cuts of the atomic density along the coordinate axes $x,y\equiv 0$, respectively. The position of the vortices have been highlighted with black circles in each corresponding phase distribution. For large positive $\mathcal{C}$, the shape of the cloud is again distorted, and regions of high density form near the edges of the trap along the $x$-axis, which can clearly be seen in panel (g). As the strength of the gauge field is decreased, the vortices arrange themselves in an ellipse, per panel (c) and (d). The final row of ground states in Fig.~\ref{fig:ell}, panels (e) and (f) correspond to a situation where there are two fewer vortices, such that the ring arrangement of (c) and (d) is broken, instead the remaining vortices are located at either edges of the cloud.                 

The nonlinear rotation gives rise to unusual vortex dynamics. For the case of \mje{rigid-body} rotation, it is always possible to define and move between a lab (LB) and co-moving (CM) frame via the transformation $|\psi_{\rm CM}\rangle=\hat{U}(t)|\psi_{\rm LB}\rangle$, yielding the Hamiltonian $\hat{H}_{\rm CM}=\hat{U}(t)\hat{H}_{\rm LB}\hat{U}(t)^{\dagger}+i\hbar\partial_{t}\hat{U}(t)\hat{U}(t)^{\dagger}$ where $\hat{U}(t)=\exp(-it\Omega({\bf r},t)\hat{L}_z/\hbar)$. Since the density-dependent gauge theory facilitates a time-dependent rotation frequency, this simple transformation \mje{gives rise to richer physics in the co-moving frame. Using the superfluid hydrodynamics equations associated with Eq.~\eqref{eqn:gpe_rot}, the transformed Hamiltonian can be shown to be 
\begin{equation}\label{eqn:hcm}
\hat{H}_{\rm CM}=\hat{H}_{\rm 2D}+\mathcal{C}t\bigg[\frac{i}{\hbar}\Omega({\bf r},t)\hat{L}_z n-\nabla\cdot(n{\bf v}_{\phi})\bigg]\hat{L}_z
\end{equation}
here the superfluid velocity is ${\bf v}_{\phi}=\hbar\nabla\phi/m$ and $\hat{H}_{\rm 2D}={\bf p}^2/2m+V({\bf r})+g_{\rm eff}n$ defines the mean-field contribution to $\hat{H}_{\rm CM}$. Equation \eqref{eqn:hcm} demonstrates} that this nonlinear system does not possess a global laboratory frame of reference. \mje{While rigid-body rotation corresponds to a global gauge symmetry of the condensate, the additional density-dependent rotation contributes a local gauge symmetry instead. Situations concerning a spatially-varying synthetic magnetic field have not been so widely studied in the literature \cite{su_2015,murray_2009}, however these systems would also lack a global laboratory frame of reference, due to the absence of global gauge symmetry.} A simple illustration of the consequence of this is presented in Fig.~\ref{fig:dyn}. Here the dynamics of two vortices are prepared such that initially they are in a non-symmetric arrangement such that $(x_{\rm v1},y_{\rm v1})=(0,a_x)$ and $(x_{\rm v2},y_{\rm v2})=(0,-2a_x)$ (panels (a,d)). Then one can see that the time evolution (panels (b,c,e,f)) causes the dynamics with $\Omega=0$ to differ depending on the value of $\mathcal{C}$. Panel (g) presents the space-time trajectories of both situations.

{\label{sec:con}\it Conclusions. } We have calculated the ground state configurations of a trapped atomic Bose-Einstein condensate in a density-dependent gauge theory. Depending on the sign and magnitude of the corresponding nonlinear rotation, different vortex configurations were obtained. When the strength of the nonlinear rotation is large and positive, almost flat-topped vortex distributions were found, whereas for large negative nonlinear rotation strengths the vortices instead arrange into ring structures. For condensates experiencing a highly anisotropic confining potential, the vortices found at large negative rotation frequencies separate into two regions at opposing edges of the trap, breaking the ring configuration. We also examined the dynamics of the vortices in this system, which are unusual due to the lack of a global laboratory reference frame. It would be interesting in the future to study nonlinear rotation in the three-dimensional situation as well as turbulent vortex dynamics.        

{\label{sec:ack}\it Acknowledgements. } MJE would like to thank Sven Bjarke Gudnason for assistance with the HPC aspects of this work and Salvatore Butera for comments on the manuscript. Numerical calculations were carried out using the TSC-computer of the ``Topological Science" project at Keio University. This work was supported by the Ministry of Education, Culture, Sports, Science (MEXT)-Supported Program for the Strategic Research Foundation at Private Universities ``Topological Science (Grant No. S1511006). This work of M.N. is also supported in part by JSPS KAKENHI Grant Numbers 16H03984, and 18H01217 and by a Grant-in-Aid for Scientific Research on Innovative Areas ``Topological Materials Science (KAKENHI Grant No. 15H05855) from MEXT of Japan.

\end{document}